\documentclass[conference]{IEEEtran}
\IEEEoverridecommandlockouts
\usepackage{cite}
\usepackage{amsmath,amssymb,amsfonts}
\usepackage{algorithmic}
\usepackage{graphicx}
\usepackage{textcomp}
\usepackage{graphicx}    % for \includegraphics
\usepackage{caption}     % for \caption inside figures
\usepackage{subcaption}
\usepackage{xcolor}
\usepackage{multirow}
\usepackage{booktabs}
\def\BibTeX{{\rm B\kern-.05em{\sc i\kern-.025em b}\kern-.08em
    T\kern-.1667em\lower.7ex\hbox{E}\kern-.125emX}}
\begin{document}

\title{Reliable Noninvasive Glucose Sensing via CNN-Based Spectroscopy\\

\thanks{This work was supported in part by the National Institute On Aging of the NIH under Award Number P30AG073105. 
The content is solely the responsibility of the authors and does not necessarily represent the official views of the NIH.}
\thanks{This work was also partially supported by the Georgia Research Alliance under Award Number GRA.25.016.KSU.01.a.}%
\thanks{All authors are with the College of Computing and Software Engineering, Kennesaw State University, Kennesaw, GA, USA. 
El Arbi Belfarsi is a Ph.D. Candidate (ebelfars@students.kennesaw.edu), 
Henry Flores is a Graduate Student (hfloresr@students.kennesaw.edu), 
and Dr. Maria Valero is an Associate Professor (mvalero2@kennesaw.edu).}

}

\author{
\IEEEauthorblockN{El Arbi Belfarsi}
\IEEEauthorblockA{
\textit{Department of Computer Science} \\
\textit{Kennesaw State University} \\
Kennesaw, GA, USA \\
ebelfars@students.kennesaw.edu}
\and
\IEEEauthorblockN{Henry Flores}
\IEEEauthorblockA{
\textit{Department of Computer Science} \\
\textit{Kennesaw State University} \\
Kennesaw, GA, USA \\
hfloresr@students.kennesaw.edu}
\and
\IEEEauthorblockN{Maria Valero}
\IEEEauthorblockA{
\textit{Department of Information Technology} \\
\textit{Kennesaw State University} \\
Kennesaw, GA, USA \\
mvalero2@kennesaw.edu}
}

\maketitle

\begin{abstract}
In this study, we present a dual‐modal AI framework based on short‐wave infrared (SWIR) spectroscopy. The first modality employs a multi‐wavelength SWIR imaging system coupled with convolutional neural networks (CNNs) to capture spatial features linked to glucose absorption. The second modality uses a compact photodiode voltage sensor and machine learning regressors (e.g., random forest) on normalized optical signals. Both approaches were evaluated on synthetic blood phantoms and skin‐mimicking materials across physiological glucose levels (70–200 mg/dL). The CNN achieved a mean absolute percentage error (MAPE) of 4.82\% at 650 nm with 100 \% Zone A coverage in the Clarke Error Grid, while the photodiode system reached 86.4\% Zone A accuracy. This framework constitutes a state-of-the-art solution that balances clinical accuracy, cost efficiency, and wearable integration, paving the way for reliable continuous non-invasive glucose monitoring.

\end{abstract}


\section{Introduction}

%\subsection{General Background}

Diabetes mellitus affects over 536 million adults worldwide (10.5\% of adults) as of 2021, and is projected to reach 783 million by 2045 \cite{Sun2022}. Maintaining blood glucose within target ranges is essential to prevent complications such as neuropathy, retinopathy, and cardiovascular disease \cite{DCCT1993}. Frequent self-monitoring of blood glucose (SMBG) via finger-prick tests remains the standard approach; however, it is invasive, painful, and can reduce patient adherence by over 63\% of the recommended levels \cite{Jacoby2010}. Continuous glucose monitoring (CGM) systems mitigate some discomfort by providing near real-time readings but still require subcutaneous sensors, carry high costs, and can suffer accuracy drift during rapid glycemic changes \cite{Alsunaidi2021, VillenaGonzales2019}.

%\subsection{Motivation for Noninvasive Glucose Monitoring}

The invasiveness and expense of current methods limit frequent monitoring, especially in resource‐constrained settings. A truly noninvasive glucometer would enhance patient compliance and quality of life by offering painless, affordable, and convenient measurement \cite{VillenaGonzales2019}. Despite decades of research, no technology has yet achieved the accuracy required to replace SMBG or CGM \cite{VillenaGonzales2019}.

%\subsection{Current Noninvasive Technologies and Limitations}

Optical methods dominate noninvasive research \cite{Alsunaidi2021}. Near-infrared (NIR) spectroscopy (750–2500 nm) can penetrate skin and interact with glucose absorption bands but suffers from weak glucose signals overlapped by water and protein peaks \cite{Yadav2015, Liu2015}. Table~\ref{tab:absorption} summarizes these overlapping absorption features \cite{peng2022use}.

\begin{table}[!htb]
\caption{NIR absorption peaks of key biological components \cite{peng2022use}.}
\label{tab:absorption}
\centering
\begin{tabular}{|c|l|l|}
\hline
\textbf{S.no} & \textbf{Component} & \textbf{Absorption Peaks (nm)} \\ \hline
1 & Glucose      & 1408, 1536, 1688, 2261 \\ \hline
2 & Water        & 1450, 1787, 1934 \\ \hline
3 & Fat          & 2299, 2342 \\ \hline
4 & Protein      & 2174, 2288 \\ \hline
\end{tabular}
\end{table}

Multivariate calibration or machine learning is required to isolate glucose signals \cite{Liu2015, Zhang2013}. Factors such as skin thickness, hydration, and temperature add variability \cite{VillenaGonzales2019}. Mid-infrared (MIR) offers stronger glucose absorbance but poor penetration \cite{Yadav2015}. Raman spectroscopy provides better specificity but weak signals and system complexity limit clinical feasibility \cite{Alsunaidi2021}. Other techniques (electromagnetic sensing, polarimetry, photoacoustic spectroscopy) face challenges in signal-to-noise ratio, tissue heterogeneity, and calibration \cite{kazi2024advancements}.

\begin{table*}[!t]
\caption{Summary of Related Work: Approach, Model, RMSE, and CEG Zone A Coverage.}
\label{tab:related_work}
\centering
\begin{tabular}{|l|l|l|c|c|}
\hline
\textbf{Reference} & \textbf{Approach}                    & \textbf{Model}                            & \textbf{RMSE (mg/dL)} & \textbf{Zone A (\%)} \\ \hline
Cistola et al.~\cite{cistola2023non} & SWIR Imaging (1500–1700 nm) & Random Forest Regression (RFR) & 10.9           & 93.1           \\ \hline
Javid et al.~\cite{javid2018mobile}        & NIR Spectroscopy (700–1100 nm) & LR / MLR                         & 14.8           & 80.0           \\ \hline
Aloraynan et al.~\cite{aloraynan2022pas}       & Mid-IR Photoacoustic            & Ensemble ML Classifiers                    & 25 mg/dL & N/A            \\ \hline
Abdolrazzaghi et al.~\cite{abdolrazzaghi2021srr} & Split-Ring Resonator (1.156 GHz) & Resonator Sensing              & N/A             & N/A            \\ \hline
Zeynali et al.~\cite{zeynali2025noninvasive}   & PPG Signals                      & ResNet34-1D                       & 19.7           & 76.6           \\ \hline
Anis and Alias~\cite{anis2021iot}             & Portable NIR (900–1600 nm, IoT)   & IoT Regression (Correlation)     & N/A             & N/A            \\ \hline
\end{tabular}
\end{table*}

The short-wave infrared (SWIR) region (700–2000 nm) shows promise: it penetrates deeper with reduced scattering and includes glucose overtone bands with less water interference, notably around 1600 nm \cite{Yadav2015, Zhang2018}. Hong \textit{et al.} demonstrated that 1500–1700 nm yields improved imaging depth and contrast in biological tissues \cite{Zhang2018}.

%\subsection{Our Contributions}

We propose an AI-driven dual‐modal framework for noninvasive glucose monitoring, validated on \textit{in vitro} samples (70–200 mg/dL):

\begin{itemize}
    \item \textbf{Imaging Mode:} Multi‐wavelength SWIR imaging using 650 nm, 808 nm, 850 nm lasers and an 850 nm LED. Deep neural networks predict glucose concentration from these images.
    \item \textbf{Voltage Mode:} SWIR sensing at 1600 nm with an LED source and an InGaAs photodiode. Voltage ratios are analyzed using linear regression (LR), multiple linear regression (MLR), and random forest (RF) models.
\end{itemize}

Both modalities employ synthetic skin and blood phantoms to better replicate physiological optical properties. Subsequent sections detail our experimental setup, data acquisition, and AI‐based analysis techniques.

\section{Related Work}

Noninvasive glucose monitoring approaches can be broadly categorized into optical imaging/spectroscopy, electromagnetic/resonator sensors, and wearable/IoT devices. Below, we review six representative studies—strictly limited to those provided—and summarize their reported performance metrics.

\subsection{Optical Imaging and Spectroscopy}

Cistola et al.~\cite{cistola2023non} employed SWIR imaging (1500–1700\,nm) on synthetic skin and blood phantoms across concentrations of 50–200\,mg/dL. Using Random Forest Regression on pixel-wise reflectance features, they achieved an RMSE of 10.9\,mg/dL with 93.1\% of predictions in Zone A and 6.9\% in Zone B of the Clarke Error Grid.

Javid et al.~\cite{javid2018mobile} proposed a smartphone-based NIR spectroscopy system (700–1100\,nm) operating in reflectance and transmittance modes. Linear and multiple linear regression were applied to five discrete glucose levels (50–250\,mg/dL), yielding an RMSE of 14.8\,mg/dL and ~80\% Zone A coverage. However, when using finer 10\,mg/dL increments, RMSE rose above 18\,mg/dL and Zone A coverage dropped below 70\%, indicating reduced linear generalization.

Aloraynan et al.~\cite{aloraynan2022pas} used mid-IR photoacoustic spectroscopy with a quantum cascade laser and artificial skin phantoms. Ensemble machine learning models achieved a sensitivity of 25 mg/dL across the physiological range, though no CEG analysis was reported.

\subsection{Electromagnetic and Resonator Sensors}

Abdolrazzaghi et al.~\cite{abdolrazzaghi2021srr} developed a 1.156\,GHz split-ring resonator with loss compensation, increasing the quality factor from 190 to 3850. In aqueous and serum-like glucose solutions (1–30\,mM), they demonstrated a linear frequency shift with a detection limit of 18\,mg/dL. However, neither RMSE nor Clarke Error Grid metrics were provided.

Zeynali et al.~\cite{zeynali2025noninvasive} applied deep learning to PPG data from the VitalDB dataset (699,000 segments from 6,368 patients). Among ResNet34-1D, VGG16-1D, and CNN-LSTM models, the best-performing model, ResNet34-1D, achieved an RMSE of 19.7\,mg/dL with 76.6\% in Zone A and a 6.4\,s inference time on TinyML hardware.

Anis and Alias~\cite{anis2021iot} introduced a handheld noninvasive glucometer using NIR spectroscopy (900–1600\,nm) with IoT-based data transmission. They reported a correlation coefficient of 0.85 against an invasive reference device, though RMSE and CEG metrics were not disclosed.

\textbf{Summary and Benchmark Selection}  
Table~\ref{tab:related_work} summarizes RMSE and Zone A results from prior studies. Among them, Cistola et al.~\cite{cistola2023non} provides the most realistic experimental setup and best overall performance. Their work serves as the primary benchmark for evaluating our dual-modal system.

\section{Methodology}\label{methodology}

\subsection{Overview}

\begin{figure}
    \centering
    \includegraphics[width=1\linewidth]{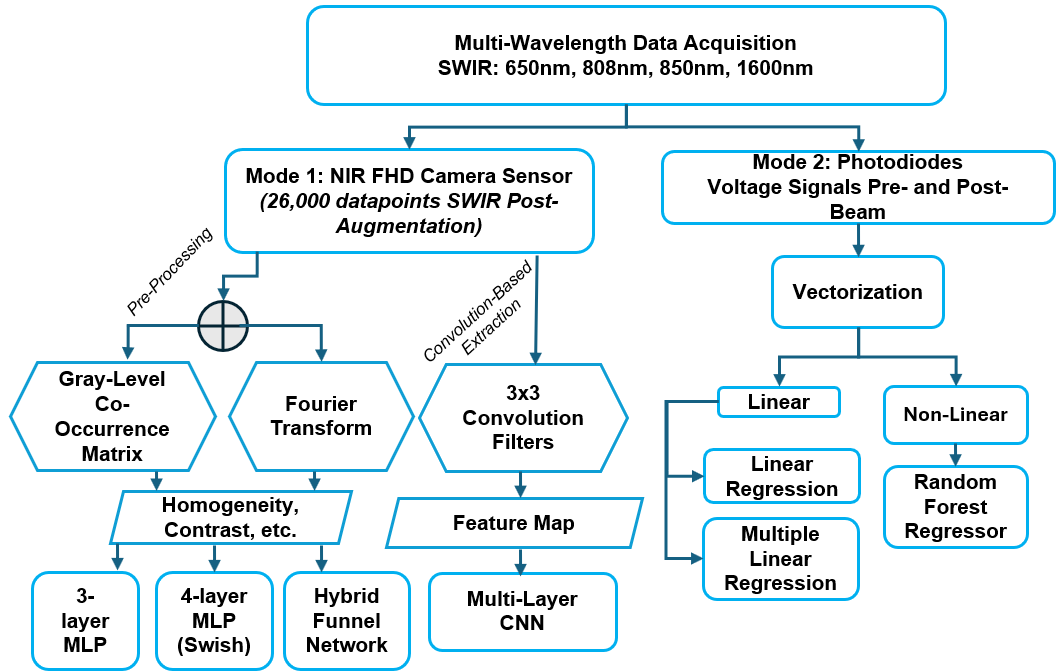}
    \caption{Block diagram representing glucose prediction approaches}
    \label{fig:block_diagram}
\end{figure}

Fig.~\ref{fig:block_diagram} showcases the overall methodology followed in the paper. In this study, we implemented two complementary modes of data acquisition to investigate non-invasive glucose estimation: (i) SWIR imaging, and (ii) photodiode-based voltage acquisition. This section details the methodology employed for each modality. 

\subsection{Mode 1: SWIR Imaging Dataset}

\subsubsection{Experimental Setup}

\begin{figure}[ht]
    \centering
    \includegraphics[width=0.7\linewidth]{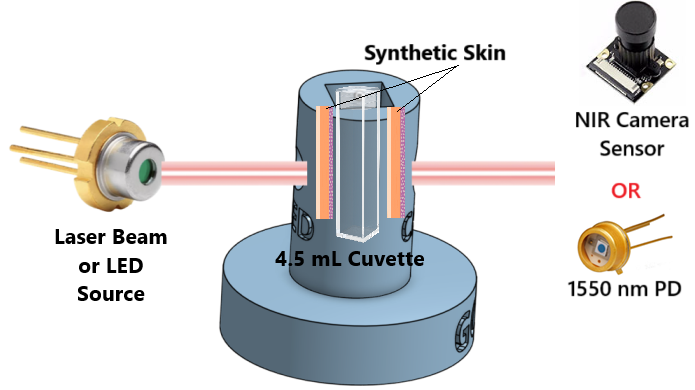}
    \caption{Schematic of the optical measurement setup using synthetic skin model with interchangeable NIR camera or photodiode sensor.}
    \label{fig:animated_design}
\end{figure}

Fig.~\ref{fig:animated_design} shows a custom 3D-printed black cuvette holder maintained consistent alignment. A 3.6 mm fully high-resolution (FHD) SWIR camera was mounted on one side, while four light sources (650 nm laser, 808 nm laser, 850 nm laser, and 850 nm LED) provided transillumination from the opposite side. Two 3 mm silicone slabs served as skin phantoms on both illumination and detection sides to mimic subcutaneous scattering.

\begin{figure}
    \centering
    \includegraphics[width=0.7\linewidth]{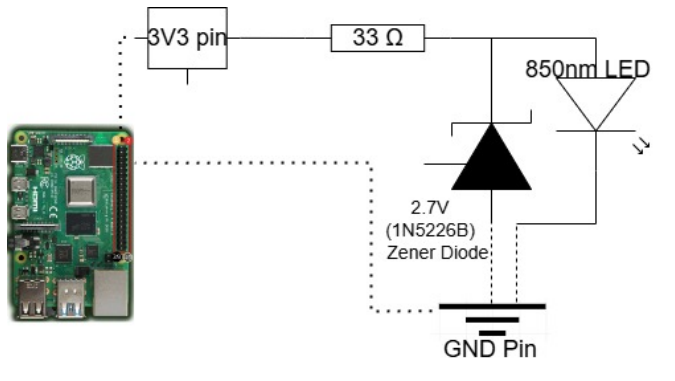}
    \caption{Circuit design for driving the LED/Laser using a 2.7 V Zener diode and 33~$\Omega$ resistor regulated by a 3.3 V supply.}
    \label{fig:circuit_design}
\end{figure}

The LED circuit (Fig.~\ref{fig:circuit_design}) consists of a 1.5 V, 50 mA LED in series with a 33~$\Omega$ resistor, powered by the Raspberry Pi’s 3V3 and GND pins. A 2.7~V, 500 mW Zener diode (1N5226B) is connected in parallel with the LED to regulate voltage and suppress transients \cite{belfarsi2025deep}.

\subsubsection{Sample Preparation}

Glucose solutions were prepared in-vitro by diluting concentrated glucose stock solutions (700–2000 mg/dL) to physiological ranges of 70–200 mg/dL, using a dilution factor of 1:10 in synthetic blood concentrate. Concentration steps were incremented by 2 mg/dL, ensuring fine-grained dataset diversity.

\subsubsection{Data Acquisition and Augmentation}

For each wavelength and glucose concentration, 10 near-infrared images were captured, resulting in a raw dataset of NIR images. To enhance model generalization and simulate real-world variances, data augmentation was applied, including:

\begin{itemize}
    \item Randomized contrast alterations,
    \item Image rotations,
    \item Application of Gaussian noise.
\end{itemize}

The final dataset consisted of approximately 6500 datapoints after augmentation, used for downstream model development and evaluation.

\subsubsection{Feature Extraction: Texture and Frequency Domain Analysis}

To enhance the model’s ability to capture complex patterns present in the SWIR images, we applied both spatial and frequency domain feature extraction techniques. These transformations are designed to extract robust descriptors that may correlate with glucose-induced optical variations \cite{haralick1973textural, gonzalez2009digital, podder2020glcm}.

\begin{figure}[ht]
    \centering
    \includegraphics[width=1\linewidth]{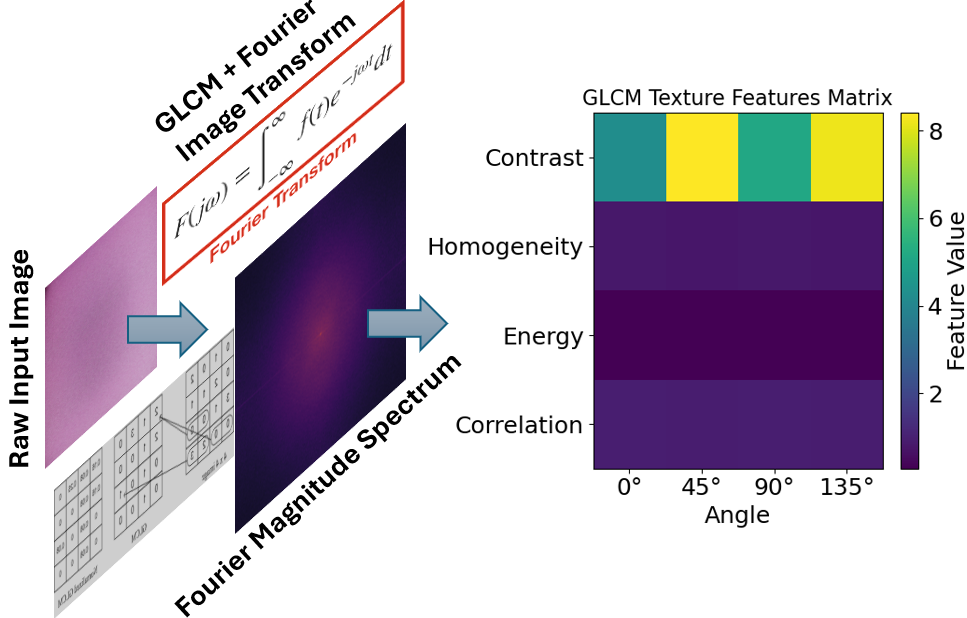}
    \caption{Feature extraction using Fourier transform and GLCM texture analysis on the input image.}
    \label{fig:prep}
\end{figure}
\paragraph{Gray Level Co-Occurrence Matrix (GLCM)}

The Gray Level Co-Occurrence Matrix (GLCM) quantifies texture by measuring how often pairs of pixel intensities occur at a specified offset and orientation. For a grayscale image $I(x,y)$ of width $W$ and height $H$, quantized to $G$ gray levels, the GLCM $P(i,j,d,\theta)$ counts the occurrences where a reference pixel with intensity $i$ has a neighboring pixel with intensity $j$ at distance $d$ and angle $\theta$:

\[
P(i,j,d,\theta) = \sum_{x=1}^{W} \sum_{y=1}^{H}
\begin{cases}
1, & \text{if } I(x,y) = i 
\\
& \text{ and } I(x+d_x,y+d_y) = j 
\\
0, & \text{otherwise}
\end{cases}
\]

where $(d_x, d_y) = (d \cos\theta, d \sin\theta)$ define the pixel offset corresponding to $\theta$.

From the GLCM, several second-order statistical texture features are extracted, including:

\begin{itemize}
    \item \textbf{Contrast:} measures local intensity variation.
    \[
    \text{Contrast} = \sum_{i,j} (i - j)^2 P(i,j)
    \]
    \item \textbf{Energy:} uniformity of texture.
    \[
    \text{Energy} = \sum_{i,j} P(i,j)^2
    \]
    \item \textbf{Homogeneity:} weights P(i,j) by inverse contrast.
    \[
    \text{Homogeneity} = \sum_{i,j} \frac{P(i,j)}{1+|i-j|}
    \]
    \item \textbf{Correlation:} statistical correlation between pixel pairs.
\end{itemize}

These features capture the micro-structural organization of the optical patterns, which may vary with glucose concentration due to changes in scattering and absorption properties \cite{podder2020glcm}.

\paragraph{Fourier Transform Features}

In parallel, the 2D Discrete Fourier Transform (DFT) captures the frequency-domain representation of the image $I(x,y)$ of size $W \times H$. The transformed spectrum $F(u,v)$ is computed as:

\[
F(u,v) = \sum_{x=0}^{W-1}\sum_{y=0}^{H-1} I(x,y) e^{-j 2\pi \left( \frac{ux}{W} + \frac{vy}{H} \right)}
\]

where $(u,v)$ are the frequency coordinates corresponding to the spatial coordinates $(x,y)$.

From the magnitude spectrum $|F(u,v)|$, we extracted:

\begin{itemize}
    \item Low-frequency energy (central concentration of energy),
    \item High-frequency content (edge and fine texture variations),
    \item Spectral entropy.
\end{itemize}

Frequency domain analysis assists in characterizing periodic patterns and subtle variations not easily observable in spatial domain alone \cite{acharya2013fourier}.

\paragraph{Feature Fusion}

The extracted GLCM and Fourier features were concatenated to form a composite feature vector, which was subsequently fed into the regression model alongside the raw image inputs. This hybrid approach aims to capitalize on both spatial texture and global frequency patterns relevant to glucose-induced optical changes. In Fig.~\ref{fig:prep}, the feature extraction pipeline illustrates the sequential computation of the Fourier magnitude spectrum and GLCM texture metrics, which jointly contribute to the model’s predictive performance.

\begin{figure*}[ht]
\centering
\includegraphics[width=0.9\textwidth]{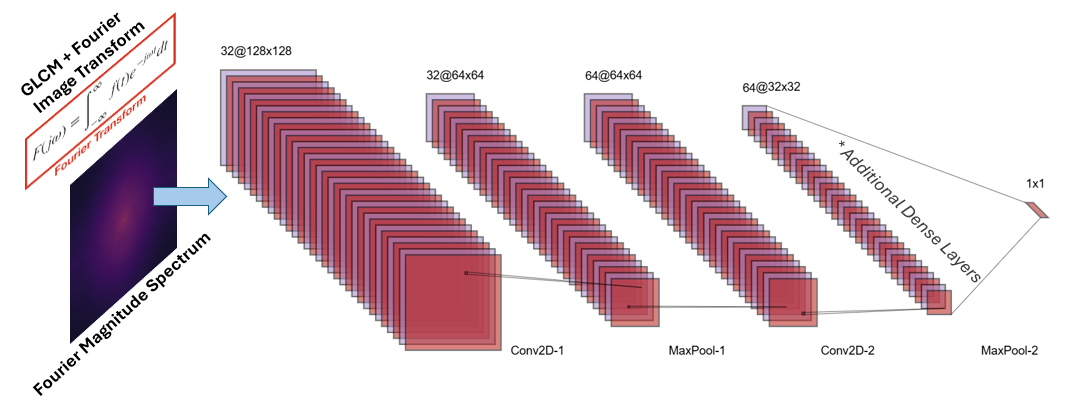}
\caption{Architecture of the feature extraction pipeline. The raw image undergoes Fourier transform followed by GLCM computation to extract texture features (contrast, homogeneity, energy, and correlation) for regression modeling.}
\label{fig:cnn_model}
\end{figure*}

\subsection{Mode 2: Photodiode Voltage Readings}

\subsubsection{Experimental Setup}

To complement the SWIR imaging modality, a simplified hardware configuration was developed by replacing the IR camera with a photodiode receiver. The updated setup consists of:

\begin{itemize}
    \item \textbf{LED Source:} A 1600 nm LED was selected as the illumination source, targeting absorption peaks of glucose molecules in the short-wave infrared (SWIR) spectrum.
    \item \textbf{Photodetector:} A 1550 nm photodiode receiver positioned opposite the LED to collect transmitted light intensity.
\end{itemize}

The rationale for selecting this configuration includes: (i) enabling more compact prototype designs suitable for mobile health applications, (ii) conducting reproducibility studies to validate findings from the imaging dataset, and (iii) exploring the underrepresented 1600 nm wavelength range, which remains relatively less investigated in the literature for noninvasive glucose sensing.

\subsubsection{Voltage Dataset Acquisition}

For each sample, baseline voltage readings were first recorded under no-beam conditions to account for system offsets. Subsequently, voltage measurements were collected both before and after the beam interaction with the glucose sample:

\begin{itemize}
    \item $V_{\text{baseline}}$: baseline voltage without illumination.
    \item $V_{\text{pre}}$: voltage recorded prior to beam activation.
    \item $V_{\text{post}}$: voltage recorded after beam transmission through the glucose solution.
\end{itemize}

The differential voltage readings captured light attenuation and scattering effects as a function of glucose concentration.

\subsubsection{Machine Learning Models}

The acquired voltage data was utilized to develop predictive regression models for glucose concentration estimation. Three models were implemented and compared:

\begin{itemize}
    \item \textbf{Linear Regression (LR):} applied as a baseline linear model.
    \item \textbf{Multiple Linear Regression (MLR):} incorporating all three voltage readings as independent variables.
    \item \textbf{Random Forest Regressor (RFR):} an ensemble learning model capable of capturing nonlinear relationships and interactions within the dataset.
\end{itemize}

\subsubsection{Feature Representation}

The feature vector for each sample was defined as:

\[
\mathbf{X} = \left[ V_{\text{baseline}}, \, V_{\text{pre}}, \, V_{\text{post}} \right]
\]

The target variable was the corresponding glucose concentration (mg/dL). Model performance was evaluated using standard regression metrics, including root mean square error (RMSE) and Clarke Error Grid analysis to assess clinical relevance.

\subsubsection{Mode 2: Photodiode-Based Model Architectures}

The photodiode-based dataset used voltage features as input, consisting of: baseline voltage ($V_{\text{baseline}}$), pre-beam voltage ($V_{\text{pre}}$), and post-beam voltage ($V_{\text{post}}$). Three regression models were evaluated, summarized in Table~\ref{tab:pd_models}.

\begin{table*}[ht]
\centering
\caption{Summary of Model Architectures}
\begin{tabular}{|p{0.7cm}|p{3cm}|p{13cm}|}
\hline
\textbf{Ref.} & \textbf{Model} & \textbf{Architecture Summary} \\
\hline
M1 & 3-layer MLP & Fully connected MLP with 3 hidden layers, \texttt{tanh} activations, and dropout (0.3) after each layer. \\
\hline
M2 & 4-layer MLP (Swish) & Fully connected MLP with 4 hidden layers, \texttt{Swish (SiLU)} activations, batch normalization. \\
\hline
M3 & Hybrid Funnel Network & Shrinking-layer architecture with Swish activations, increasing $L_2$ regularization depth-wise, and dropout. \\
\hline
M4 & Region-Based CNN & Shallow CNN with max-pooling and dropout, followed by dense layers for regression; see Fig.~\ref{fig:cnn_model}. \\
\hline
\end{tabular}
\label{tab:model_architectures}
\end{table*}

\begin{table*}[ht]
\centering
\caption{Summary of Photodiode-Based Model Architectures}
\begin{tabular}{|p{0.5cm}|p{3.3cm}|p{12cm}|}
\hline
\textbf{Ref.} & \textbf{Model} & \textbf{Architecture Summary} \\
\hline
LR & Linear Regression & Ordinary least squares using a single voltage feature to predict glucose concentration. \\
\hline
MLR & Multiple Linear Regression & Linear model using $V_{\text{baseline}}$, $V_{\text{pre}}$, and $V_{\text{post}}$ as predictors. \\
\hline
RFR & Random Forest Regressor & Tree ensemble using all voltage features with bagging and random feature selection per split. \\
\hline
\end{tabular}
\label{tab:pd_models}
\end{table*}

\section{Implementation Details}
\subsection{SWIR Imaging Operating Mode}

\subsubsection{Region-Based CNN Model Training Protocol}
Following the evaluation of multiple models (M1 through M4 shown in Table~\ref{tab:model_architectures}), we focus in this section on the best-performing architecture: the Region-Based CNN (Model M4). The full implementation was performed using Python with TensorFlow and Scikit-Learn libraries. The dataset was first filtered by wavelength, and separate models were trained for each wavelength independently.

A global image resizing step was applied to standardize all SWIR images to $128 \times 128$ pixels, followed by normalization of pixel intensities to the $[0, 1]$ range. For each wavelength subset, images and corresponding glucose concentration labels were extracted for supervised learning.

As per cross-validation, we ensured reliable evaluation by applying a single-shot holdout strategy with a 70:30 train-test split, randomly partitioned with a fixed seed for reproducibility ($\texttt{random\_state}=42$). All reported results are based on the unseen test set performance.
\subsubsection{Experimental Workflow}

The model was trained using the Adam optimizer with default learning rate settings. The loss function was Mean Squared Error (MSE), with Mean Absolute Error (MAE) monitored as an auxiliary metric. Each model was trained for 50 epochs with a batch size of 32. Early stopping was not applied in order to ensure full convergence across all datasets.

\subsection{Voltage Reading Mode}

\subsubsection{Random Forest Regression Details}

In the context of the photodiode-based voltage recordings, we evaluated multiple regression algorithms and determined that Random Forest Regression (RFR) consistently provided the most accurate and stable performance for glucose concentration estimation. Its ability to capture nonlinear relationships inherent in the processed voltage signals made it the most suitable choice for this modality. As such, we present its implementation details below.

\subsubsection{Model Configuration}

The RFR model was implemented using the \texttt{scikit-learn} library in Python. Following parameter tuning, the number of estimators (\texttt{n\_estimators}) was set to 100, and the maximum tree depth (\texttt{max\_depth}) was limited to 15 to maintain generalization while preventing overfitting. The input features consisted of the normalized ratio of post-beam to pre-beam voltages obtained from the photodiode sensors. Feature scaling was performed using min-max normalization to ensure consistent input ranges across the ensemble learners.

\section{Results}

\subsection{Overview}

Table~\ref{tab:full_results_summary} presents a full comparative summary of model performances across different wavelengths and preprocessing strategies. The models are identified according to the architecture codes introduced earlier. All reported metrics were computed on the unseen test sets using the evaluation framework described in Section \ref{methodology}.

\begin{table*}[ht]
\centering
\caption{Summary of Model Performance Across Modalities, Preprocessing Strategies, and Wavelengths}
\begin{tabular}{p{2.5cm} p{2.0cm} p{3.6cm} p{1.5cm} p{1.2cm} p{1.2cm} p{1.2cm}}
\toprule
\textbf{Mode} & \textbf{Pre-Processing} & \textbf{Model} & \textbf{Wavelength} & \textbf{RMSE} & \textbf{MAE} & \textbf{MAPE} \\
\midrule

\multirow{12}{*}{\textbf{SWIR Imaging}}

& \multirow{12}{*}{\parbox{2.0cm}{Gray-Level\\ Co-Occurrence\\ Matrix \&\\ Fourier Domain}} 
& 3-Layer MLP (\textbf{M1}) & 650 & 15.32 & 12.20 & 10.5\% \\
& & & 808 & 19.71 & 15.70 & 13.64\% \\
& & & 850$^{\text{avg}}$ & 17.26 & 13.80 & 11.8\% \\
\cmidrule(lr){3-7}

& & 4-Layer MLP (\textbf{M2}) & 650 & 13.83 & 10.30 & 9.55\% \\
& & & 808 & 16.70 & 11.70 & 10.56\% \\
& & & 850$^{\text{avg}}$ & 18.62 & 14.85 & 13.05\% \\
\cmidrule(lr){3-7}

& & Hybrid Funnel (\textbf{M3}) & 650 & 8.36 & 6.30 & 5.47\% \\
& & & 808 & 19.46 & 15.59 & 13.21\% \\
& & & 850$^{\text{avg}}$ & 11.27 & 8.22 & 7.31\% \\
\cmidrule(lr){2-7}

& \multirow{2}{*}{\parbox{2.0cm}{Convolution-Based Features Extraction}}
& Region-Based CNN (\textbf{M4}) & 650 & 8.29 & 6.31 & \textbf{4.82}\% \\
& & & 808 & 15.19 & 12.10 & 9.42\% \\
& & & 850$^{\text{avg}}$ & 12.33 & 9.32 & 7.52\% \\
\midrule

\multirow{4}{*}{\textbf{Voltage Readings}}
& \multirow{2}{*}{\parbox{2.0cm}{Post-Beam to \\Pre-Beam Ratio}}
& Linear Regression (\textbf{LR}) & 1600 & - &  - &  - \\
& & Multiple Linear Regression (\textbf{MLR}) & 1600 & - &  - &  - \\
& & Random Forest Regressor (\textbf{RFR}) & 1600  & 17.62 & 14.05 & 14.07\% \\
\bottomrule

\end{tabular}
\label{tab:full_results_summary}
\end{table*}

The voltage-based photodiode modality achieves its lowest error when using nonlinear models: Random Forest Regressor (RFR) reached a MAPE of 14.1\% after post-/pre-beam normalization, outperforming simple Linear Regression (LR) and Multiple LR (MLR), which fail to capture the optical signal’s nonlinear dependencies~\cite{javid2018noninvasive}. Therefore, further analysis is required to verify reproducibility. 

In contrast, SWIR imaging with convolutional feature extraction via a Region-Based CNN achieved a MAPE of 4.82\% at 650~nm, demonstrating superior spatial information use. However, CNNs demand substantial computational resources and hardware, potentially limiting deployment in low-power wearables.

Photodiode systems, while less accurate, remain attractive for ambulatory monitoring due to their small size, low cost, and minimal power requirements when coupled with robust nonlinear regressors.

\subsection{Wavelength Optimality Analysis}

\begin{figure}[ht]
    \centering
    \includegraphics[width=0.8\linewidth]{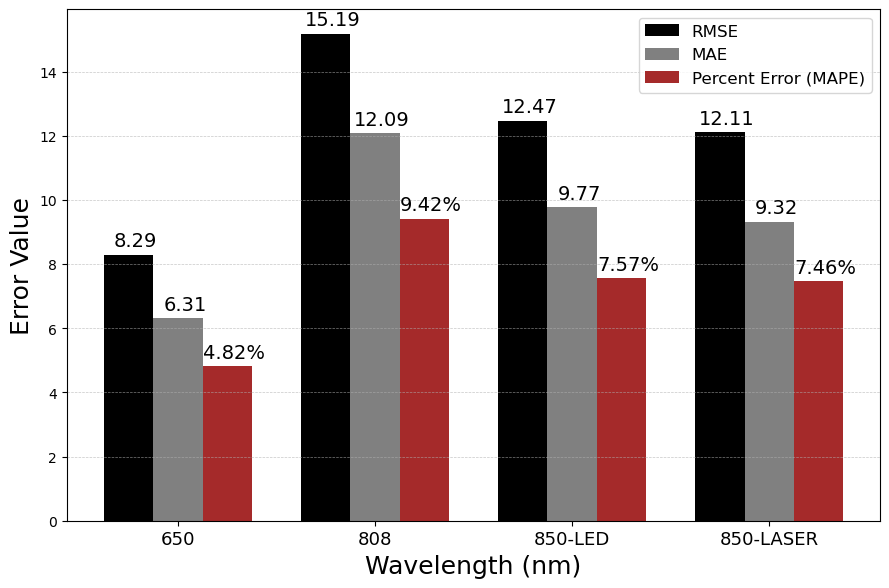}
    \caption{Error metrics (RMSE, MAE, MAPE) for CNN-based glucose prediction at different wavelengths.}
    \label{fig:wavelength-optimality}
\end{figure}

To determine the most effective wavelength for CNN-based glucose prediction, we evaluated performance metrics across four illumination conditions: 650 nm Laser, 808 nm Laser, 850 nm LED, and 850 nm Laser. Fig.~\ref{fig:wavelength-optimality} summarizes the Root Mean Squared Error (RMSE), Mean Absolute Error (MAE), and Mean Absolute Percentage Error (MAPE) for each wavelength.

As shown in Fig.~\ref{fig:wavelength-optimality}, the lowest RMSE, MAE, and MAPE values at 650 nm indicate that this wavelength provides the highest signal‐to‐noise ratio for imaging glucose-dependent spectral changes in our setup. Both 850 nm conditions (LED and Laser) achieve intermediate error rates; however, the 850 nm Laser slightly outperforms the 850 nm LED in terms of MAE and MAPE, suggesting that coherent illumination at 850 nm yields marginally better feature extraction over 808 nm for our CNN.

Overall, these results demonstrate that 650 nm Laser illumination is optimal for CNN‐based in vitro glucose prediction under our experimental conditions. Consequently, we adopt 650 nm as the primary wavelength for subsequent model refinement and integration into the final glucometer design.

\begin{figure*}[ht]
  \centering
  % Four CNN subfigures in one row
  \begin{subfigure}[b]{0.24\textwidth}
    \centering
    \includegraphics[width=\linewidth]{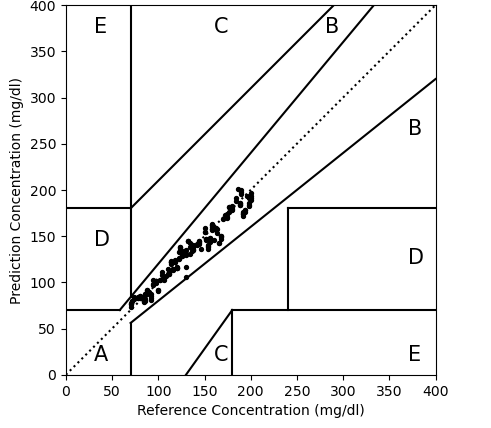}
    \caption{650 nm Laser (CNN)}
    \label{fig:ceg-650-laser}
  \end{subfigure}\hfill
  \begin{subfigure}[b]{0.24\textwidth}
    \centering
    \includegraphics[width=\linewidth]{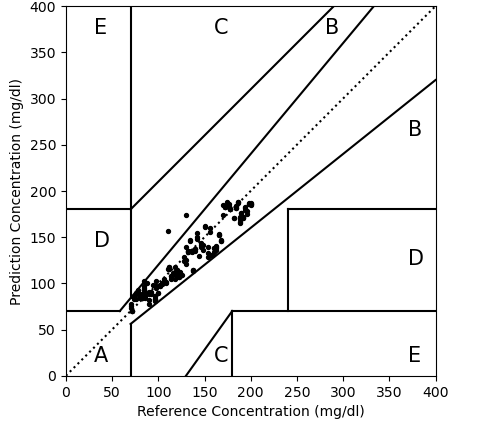}
    \caption{850 nm LED (CNN)}
    \label{fig:ceg-850-led}
  \end{subfigure}\hfill
  \begin{subfigure}[b]{0.24\textwidth}
    \centering
    \includegraphics[width=\linewidth]{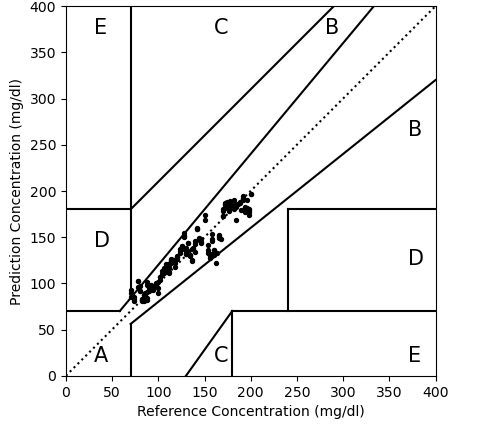}
    \caption{850 nm Laser (CNN)}
    \label{fig:ceg-850-laser}
  \end{subfigure}\hfill
  \begin{subfigure}[b]{0.24\textwidth}
    \centering
    \includegraphics[width=\linewidth]{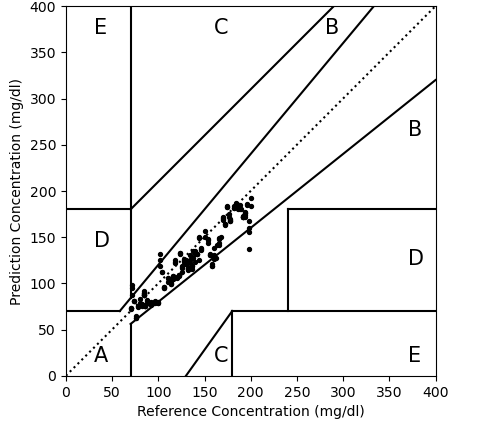}
    \caption{808 nm Laser (CNN)}
    \label{fig:ceg-808-laser}
  \end{subfigure}

  \caption{Clarke Error Grid results for the CNN model at (a) 650 nm Laser, (b) 850 nm LED, (c) 850 nm Laser, and (d) 808 nm Laser.}
  \label{fig:ceg-cnns}
\end{figure*}

\subsection{Clarke Error Grid Analysis}

The Clarke Error Grid (CEG) was employed to provide a clinically relevant evaluation of the glucose prediction accuracy across all experimental conditions. We present the results for each modality individually (Figs.~\ref{fig:ceg-650-laser}--\ref{fig:ceg-1600-rfr}), followed by a comparison with prior studies.

\subsubsection{CNN-Based Image Mode}

CNNs were trained independently on image data under different wavelength and illumination settings. Figs.~\ref{fig:ceg-650-laser}–\ref{fig:ceg-808-laser} show the CEG plots:

\begin{itemize}
    \item \textbf{650\,nm Laser}: 100.0\% in Zone A.
    \item \textbf{850\,nm LED}: 98.5\% Zone A, 1.5\% Zone B.
    \item \textbf{850\,nm Laser}: 94.2\% Zone A, 4.2\% B, 1.6\% D.
    \item \textbf{808\,nm Laser}: 91.5\% Zone A, 8.5\% Zone B.
\end{itemize}

Linear baselines from Javid et al.~\cite{javid2018noninvasive} failed to generalize under our 2\,mg/dL resolution. We attribute this to their coarser sampling, which masked the limits of linear models under finer-grained conditions.

\subsubsection{Photodiode-Based Voltage Mode}

\begin{figure}[ht]
  \centering
  \includegraphics[width=0.65\linewidth]{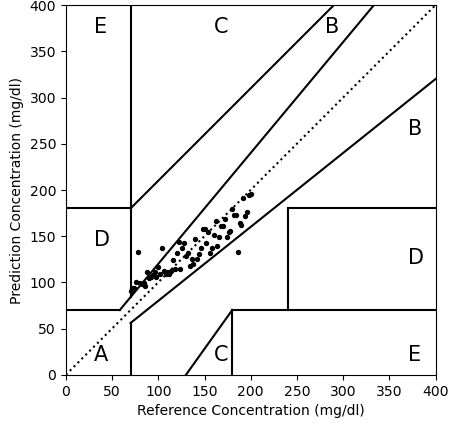}
  \caption{Clarke Error Grid for photodiode-based voltage mode (1600\,nm LED, RFR) – 86.4\% Zone A.}
  \label{fig:ceg-1600-rfr}
\end{figure}

Fig.~\ref{fig:ceg-1600-rfr} shows the CEG for the RFR model on 1600\,nm photodiode voltage ratios:

\begin{itemize}
    \item \textbf{RFR (1600\,nm LED)}: 86.4\% Zone A, 13.6\% Zone B.
\end{itemize}

While trailing CNN performance, this mode remains clinically viable and offers a compact, cost-effective alternative for real-time glucose monitoring.

\subsubsection{Comparison with Existing Work}

Our best Zone A results (100.0\% at 650\,nm, 98.5\% at 850\,nm) exceed those of Cistola et al.~\cite{cistola2023non}, despite their simpler aqueous glucose setup. By incorporating synthetic skin and chromatic blood mimics, our system better approximates real-world conditions and confirms the robustness of both CNN and RFR models under clinically relevant constraints.

\section{Conclusion and Future Work}

We have demonstrated a dual‐modal AI framework for noninvasive glucose monitoring using SWIR spectroscopy. A CNN‐based imaging system achieved a MAPE of 4.82\% at 650 nm with 100\% Zone A coverage on the Clarke Error Grid, while a compact photodiode‐voltage sensor reached 86.4\% Zone A accuracy. Together, these modalities offer a balance of high clinical fidelity and practical advantages in cost, size, and power consumption. By harnessing both spatial and spectral information, our approach advances beyond traditional invasive methods and prior optical techniques.

Looking ahead, we plan to validate these models in vivo across diverse subjects and real‐world settings, including motion, temperature, and perspiration variability. We will also explore adaptive calibration to account for skin tone, hydration, and tissue heterogeneity. On the engineering side, we aim to streamline the CNN pipeline for edge devices, reduce inference latency, and investigate hybrid modeling strategies that fuse imaging and voltage data at runtime. These efforts will bring us closer to a robust, wearable continuous glucose monitor suitable for daily use.

\bibliographystyle{IEEEtran}
\bibliography{references}

\end{document}